%% file: main.tex
\pdfoutput=1

\documentclass[12pt,a4paper]{article}

\usepackage[colorinlistoftodos]{todonotes}

\usepackage{ifthen} 
\newboolean{pdflatex}
\setboolean{pdflatex}{true} 

\newboolean{articletitles}
\setboolean{articletitles}{true} 

\newboolean{uprightparticles}
\setboolean{uprightparticles}{false} 

\newboolean{inbibliography}
\setboolean{inbibliography}{false} 

\def\paperauthors{LHCb collaboration} 
\def\paperasciititle{Search for the lepton-flavour-violating decays Bs -> tau mu and Bd -> tau mu} 
\def\papertitle{Search for the lepton-flavour-violating decays \BsToTauMu and \BdToTauMu}
\def\paperkeywords{{High Energy Physics}, {LHCb}} 
\def\papercopyright{\the\year\ CERN for the benefit of the LHCb collaboration} 
\def\paperlicence{CC-BY-4.0 licence}
\def\paperlicenceurl{https://creativecommons.org/licenses/by/4.0/}

\input{preamble}
\usepackage{longtable} 

\begin{document}


\input{title-LHCb-PAPER}


\renewcommand{\thefootnote}{\arabic{footnote}}
\setcounter{footnote}{0}



\pagestyle{plain} 
\setcounter{page}{1}
\pagenumbering{arabic}


\input{content}
\input{acknowledgements}

\addcontentsline{toc}{section}{References}
\setboolean{inbibliography}{true}
\bibliographystyle{LHCb}
\bibliography{main,standard,LHCb-PAPER,LHCb-CONF,LHCb-DP,LHCb-TDR}
\clearpage
\input{supplemental}


\clearpage 
\input{LHCb_Authorship_10-Mar-2019.tex}

\end{document}

%% file: preamble.tex
\usepackage[top=1in, bottom=1.25in, left=1in, right=1in]{geometry}

\usepackage{microtype}
\usepackage{lineno}  
\usepackage{xspace} 
\usepackage{caption} 

\usepackage{graphicx}  
\usepackage{color}
\usepackage{colortbl}
\graphicspath{{./}{./figs/}{./figs/Supplementary/}} 
\DeclareGraphicsExtensions{.pdf,.PDF,png,.PNG}

\usepackage{amsmath} 
\usepackage{amssymb}
\usepackage{amsfonts}
\usepackage{upgreek} 

\newcommand*\patchAmsMathEnvironmentForLineno[1]{%
\expandafter\let\csname old#1\expandafter\endcsname\csname #1\endcsname
\expandafter\let\csname oldend#1\expandafter\endcsname\csname
end#1\endcsname
 \renewenvironment{#1}%
   {\linenomath\csname old#1\endcsname}%
   {\csname oldend#1\endcsname\endlinenomath}%
}
\newcommand*\patchBothAmsMathEnvironmentsForLineno[1]{%
  \patchAmsMathEnvironmentForLineno{#1}%
  \patchAmsMathEnvironmentForLineno{#1*}%
}
\AtBeginDocument{%
\patchBothAmsMathEnvironmentsForLineno{equation}%
\patchBothAmsMathEnvironmentsForLineno{align}%
\patchBothAmsMathEnvironmentsForLineno{flalign}%
\patchBothAmsMathEnvironmentsForLineno{alignat}%
\patchBothAmsMathEnvironmentsForLineno{gather}%
\patchBothAmsMathEnvironmentsForLineno{multline}%
\patchBothAmsMathEnvironmentsForLineno{eqnarray}%
}

\usepackage{hyperxmp}

\usepackage[pdftex,
            pdfauthor={\paperauthors},
            pdftitle={\paperasciititle},
            pdfkeywords={\paperkeywords},
            pdfcopyright={Copyright (C) \papercopyright},
            pdflicenseurl={\paperlicenceurl}]{hyperref}

\usepackage[all]{hypcap} 

\input{lhcb-symbols-def} 
\input{paper-symbols-def}

\usepackage{cite} 
\usepackage{mciteplus}

%% file: lhcb-symbols-def.tex
\usepackage{xspace} 
\usepackage{upgreek}

\def\lhcb   {\mbox{LHCb}\xspace}

\def\babar  {\mbox{BaBar}\xspace}

\def\MagUp {\mbox{\em Mag\kern -0.05em Up}\xspace}

\ifthenelse{\boolean{uprightparticles}}%
{

 \def\Pmu         {\ensuremath{\upmu}\xspace}                 
 \def\Pnu         {\ensuremath{\upnu}\xspace}                 
                  
 \def\Ppi         {\ensuremath{\uppi}\xspace}

 \def\Ptau        {\ensuremath{\uptau}\xspace}

 \def\Ppsi        {\ensuremath{\uppsi}\xspace}

 \def\PDelta      {\ensuremath{\Delta}\xspace}                 
 \def\PXi         {\ensuremath{\Xi}\xspace}                 
 \def\PLambda     {\ensuremath{\Lambda}\xspace}                 
 \def\PSigma      {\ensuremath{\Sigma}\xspace}                 
 \def\POmega      {\ensuremath{\Omega}\xspace}                 
 \def\PUpsilon    {\ensuremath{\Upsilon}\xspace}

 \def\PB      {\ensuremath{\mathrm{B}}\xspace}                 
                  
 \def\PD      {\ensuremath{\mathrm{D}}\xspace}

 \def\PJ      {\ensuremath{\mathrm{J}}\xspace}                 
 \def\PK      {\ensuremath{\mathrm{K}}\xspace}

 \def\PZ      {\ensuremath{\mathrm{Z}}\xspace}                 
                  
 \def\Pb      {\ensuremath{\mathrm{b}}\xspace}                 
 \def\Pc      {\ensuremath{\mathrm{c}}\xspace}

 \def\Pi      {\ensuremath{\mathrm{i}}\xspace}

 \def\Ps      {\ensuremath{\mathrm{s}}\xspace}

}
{

 \def\Pmu         {\ensuremath{\mu}\xspace}                 
 \def\Pnu         {\ensuremath{\nu}\xspace}                 
                  
 \def\Ppi         {\ensuremath{\pi}\xspace}

 \def\Ptau        {\ensuremath{\tau}\xspace}

 \def\Ppsi        {\ensuremath{\psi}\xspace}                 
                  
 \mathchardef\PDelta="7101
 \mathchardef\PXi="7104
 \mathchardef\PLambda="7103
 \mathchardef\PSigma="7106
 \mathchardef\POmega="710A
 \mathchardef\PUpsilon="7107
                  
 \def\PB      {\ensuremath{B}\xspace}                 
                  
 \def\PD      {\ensuremath{D}\xspace}

 \def\PJ      {\ensuremath{J}\xspace}                 
 \def\PK      {\ensuremath{K}\xspace}

 \def\PZ      {\ensuremath{Z}\xspace}                 
                  
 \def\Pb      {\ensuremath{b}\xspace}                 
 \def\Pc      {\ensuremath{c}\xspace}

 \def\Pi      {\ensuremath{i}\xspace}

 \def\Ps      {\ensuremath{s}\xspace}

}

\makeatletter
\ifcase \@ptsize \relax
  \newcommand{\miniscule}{\@setfontsize\miniscule{4}{5}}
\or
  \newcommand{\miniscule}{\@setfontsize\miniscule{5}{6}}
\or
  \newcommand{\miniscule}{\@setfontsize\miniscule{5}{6}}
\fi
\makeatother

\DeclareRobustCommand{\optbar}[1]{\shortstack{{\miniscule (\rule[.5ex]{1.25em}{.18mm})}
  \\ [-.7ex] $#1$}}

\def\muon       {{\ensuremath{\Pmu}}\xspace}
\def\mup        {{\ensuremath{\Pmu^+}}\xspace}
\def\mun        {{\ensuremath{\Pmu^-}}\xspace} 
 
\def\mump       {{\ensuremath{\Pmu^\mp}}\xspace}

\def\tauon      {{\ensuremath{\Ptau}}\xspace}

\def\taum       {{\ensuremath{\Ptau^-}}\xspace}
\def\taupm      {{\ensuremath{\Ptau^\pm}}\xspace}

\def\neu        {{\ensuremath{\Pnu}}\xspace}
\def\neub       {{\ensuremath{\overline{\Pnu}}}\xspace}
\def\neum       {{\ensuremath{\neu_\mu}}\xspace}
\def\neumb      {{\ensuremath{\neub_\mu}}\xspace}
\def\neut       {{\ensuremath{\neu_\tau}}\xspace}

\def\Z      {{\ensuremath{\PZ}}\xspace}

\def\squark    {{\ensuremath{\Ps}}\xspace}

\def\cquark    {{\ensuremath{\Pc}}\xspace}

\def\bquark    {{\ensuremath{\Pb}}\xspace}

\def\pion   {{\ensuremath{\Ppi}}\xspace}
\def\piz    {{\ensuremath{\pion^0}}\xspace}
\def\pip    {{\ensuremath{\pion^+}}\xspace}
\def\pim    {{\ensuremath{\pion^-}}\xspace}
\def\pipm   {{\ensuremath{\pion^\pm}}\xspace}
\def\pimp   {{\ensuremath{\pion^\mp}}\xspace}

\def\kaon    {{\ensuremath{\PK}}\xspace}
  \def\Kbar    {{\kern 0.2em\overline{\kern -0.2em \PK}{}}\xspace}

\def\KorKbar    {\kern 0.18em\optbar{\kern -0.18em K}{}\xspace}

\def\Kp      {{\ensuremath{\kaon^+}}\xspace}

  \def\Dbar    {{\kern 0.2em\overline{\kern -0.2em \PD}{}}\xspace}
\def\D       {{\ensuremath{\PD}}\xspace}

\def\DorDbar    {\kern 0.18em\optbar{\kern -0.18em D}{}\xspace}

\def\Dp      {{\ensuremath{\D^+}}\xspace}
\def\Dm      {{\ensuremath{\D^-}}\xspace}

\def\B       {{\ensuremath{\PB}}\xspace}
\def\Bbar    {{\ensuremath{\kern 0.18em\overline{\kern -0.18em \PB}{}}}\xspace}

\def\BorBbar    {\kern 0.18em\optbar{\kern -0.18em B}{}\xspace}
\def\Bz      {{\ensuremath{\B^0}}\xspace}

\def\Bu      {{\ensuremath{\B^+}}\xspace}

\def\Bp      {{\ensuremath{\Bu}}\xspace}

\def\Bd      {{\ensuremath{\B^0}}\xspace}
\def\Bs      {{\ensuremath{\B^0_\squark}}\xspace}

\def\BdorBs  {{\ensuremath{\B^0_{(\squark)}}}\xspace}

\def\jpsi     {{\ensuremath{{\PJ\mskip -3mu/\mskip -2mu\Ppsi\mskip 2mu}}}\xspace}

\def\Y#1S{\ensuremath{\PUpsilon{(#1S)}}\xspace}

\def\Lbar        {{\ensuremath{\kern 0.1em\overline{\kern -0.1em\PLambda}}}\xspace}
\def\LorLbar     {\kern 0.18em\optbar{\kern -0.18em \PLambda}{}\xspace}

\def\BF         {{\ensuremath{\mathcal{B}}}\xspace}

\def\BR         {\BF}
\newcommand{\decay}[2]{\mbox{\ensuremath{#1\!\to #2}}\xspace}         

\def\to                 {\ensuremath{\rightarrow}\xspace}

\def\AT#1     {\ensuremath{A_{\mathrm{T}}^{#1}}\xspace}           

\def\C#1      {\ensuremath{\mathcal{C}_{#1}}\xspace}                       
\def\Cp#1     {\ensuremath{\mathcal{C}_{#1}^{'}}\xspace}                    
\def\Ceff#1   {\ensuremath{\mathcal{C}_{#1}^{\mathrm{(eff)}}}\xspace}        
\def\Cpeff#1  {\ensuremath{\mathcal{C}_{#1}^{'\mathrm{(eff)}}}\xspace}       
\def\Ope#1    {\ensuremath{\mathcal{O}_{#1}}\xspace}                       
\def\Opep#1   {\ensuremath{\mathcal{O}_{#1}^{'}}\xspace}                    

\newcommand{\tev}{\ifthenelse{\boolean{inbibliography}}{\ensuremath{~T\kern -0.05em eV}}{\ensuremath{\mathrm{\,Te\kern -0.1em V}}}\xspace}
\newcommand{\gev}{\ensuremath{\mathrm{\,Ge\kern -0.1em V}}\xspace}
\newcommand{\mev}{\ensuremath{\mathrm{\,Me\kern -0.1em V}}\xspace}
\newcommand{\kev}{\ensuremath{\mathrm{\,ke\kern -0.1em V}}\xspace}
\newcommand{\ev}{\ensuremath{\mathrm{\,e\kern -0.1em V}}\xspace}
\newcommand{\gevc}{\ensuremath{{\mathrm{\,Ge\kern -0.1em V\!/}c}}\xspace}
\newcommand{\mevc}{\ensuremath{{\mathrm{\,Me\kern -0.1em V\!/}c}}\xspace}
\newcommand{\gevcc}{\ensuremath{{\mathrm{\,Ge\kern -0.1em V\!/}c^2}}\xspace}
\newcommand{\gevgevcccc}{\ensuremath{{\mathrm{\,Ge\kern -0.1em V^2\!/}c^4}}\xspace}
\newcommand{\mevcc}{\ensuremath{{\mathrm{\,Me\kern -0.1em V\!/}c^2}}\xspace}

\def\mum  {\ensuremath{{\,\upmu\mathrm{m}}}\xspace}

\def\invfb   {\ensuremath{\mbox{\,fb}^{-1}}\xspace}

\def\gsim{{~\raise.15em\hbox{$>$}\kern-.85em
          \lower.35em\hbox{$\sim$}~}\xspace}
\def\lsim{{~\raise.15em\hbox{$<$}\kern-.85em
          \lower.35em\hbox{$\sim$}~}\xspace}

\def\pt         {\ensuremath{p_{\mathrm{T}}}\xspace}
\def\ptot       {\ensuremath{p}\xspace}

\def\evtgen     {\mbox{\textsc{EvtGen}}\xspace}

\def\geant      {\mbox{\textsc{Geant4}}\xspace}

\def\photos     {\mbox{\textsc{Photos}}\xspace}

\def\pythia     {\mbox{\textsc{Pythia}}\xspace}

\def\tell1  {TELL1\xspace}
\def\ukl1   {UKL1\xspace}



%% file: paper-symbols-def.tex
\newcommand{\subdecay}[2]{\ensuremath{#1(\!\to #2)}\xspace}

\def\mump   {{\ensuremath{\muon^\mp}}\xspace}
\def\taupm   {{\ensuremath{\tauon^\pm}}\xspace}

\def\Ddorsm     {{\ensuremath{\D^-_{(\squark)}}}\xspace}
\def\aonem {{\ensuremath{a_{1}(1260)^{-}}}\xspace}

\def\tauppp{\decay{\taum}{\pim\pip\pim\neut}}
\def\tauppppiz{\decay{\taum}{\pim\pip\pim\piz\neut}}

\def\Dkpp{\decay{\Dm}{\Kp\pim\pim}}
\def\BdToTauMu   {\decay{\Bd}{\taupm\mump}}
\def\BsToTauMu   {\decay{\Bs}{\taupm\mump}}
\def\BdorsToTauMu   {\decay{\BdorBs}{\taupm\mump}}

\def\Bstaumu{\decay{\Bs}{\subdecay{\taupm}{\pipm\pimp\pipm\neut}\mump}}

\def\Bstaumupi0{\decay{\Bs}{\subdecay{\taupm}{\pipm\pimp\pipm\piz\neu}\mump}}
\def\Bdtaumupi0{\decay{\Bd}{\subdecay{\taupm}{\pipm\pimp\pipm\piz\neu}\mump}}
\def\BToDpi    {\decay{\Bz}{\Dm\pip}}
\def\BDpi{\decay{\Bz}{\subdecay{\Dm}{\Kp\pim\pim}\pip}}
\def\BToJpsiK{\decay{\Bp}{\subdecay{\jpsi}{\mup\mun}\Kp}}
\def\tauola     {\mbox{\textsc{Tauola}}\xspace}

\def\MB{{\ensuremath{M_{B}}}\xspace}



%% file: title-LHCb-PAPER.tex

\begin{titlepage}
\pagenumbering{roman}

\vspace*{-1.5cm}
\centerline{\large EUROPEAN ORGANIZATION FOR NUCLEAR RESEARCH (CERN)}
\vspace*{1.5cm}
\noindent
\begin{tabular*}{\linewidth}{lc@{\extracolsep{\fill}}r@{\extracolsep{0pt}}}
\ifthenelse{\boolean{pdflatex}}
{\vspace*{-1.5cm}\mbox{\!\!\!\includegraphics[width=.14\textwidth]{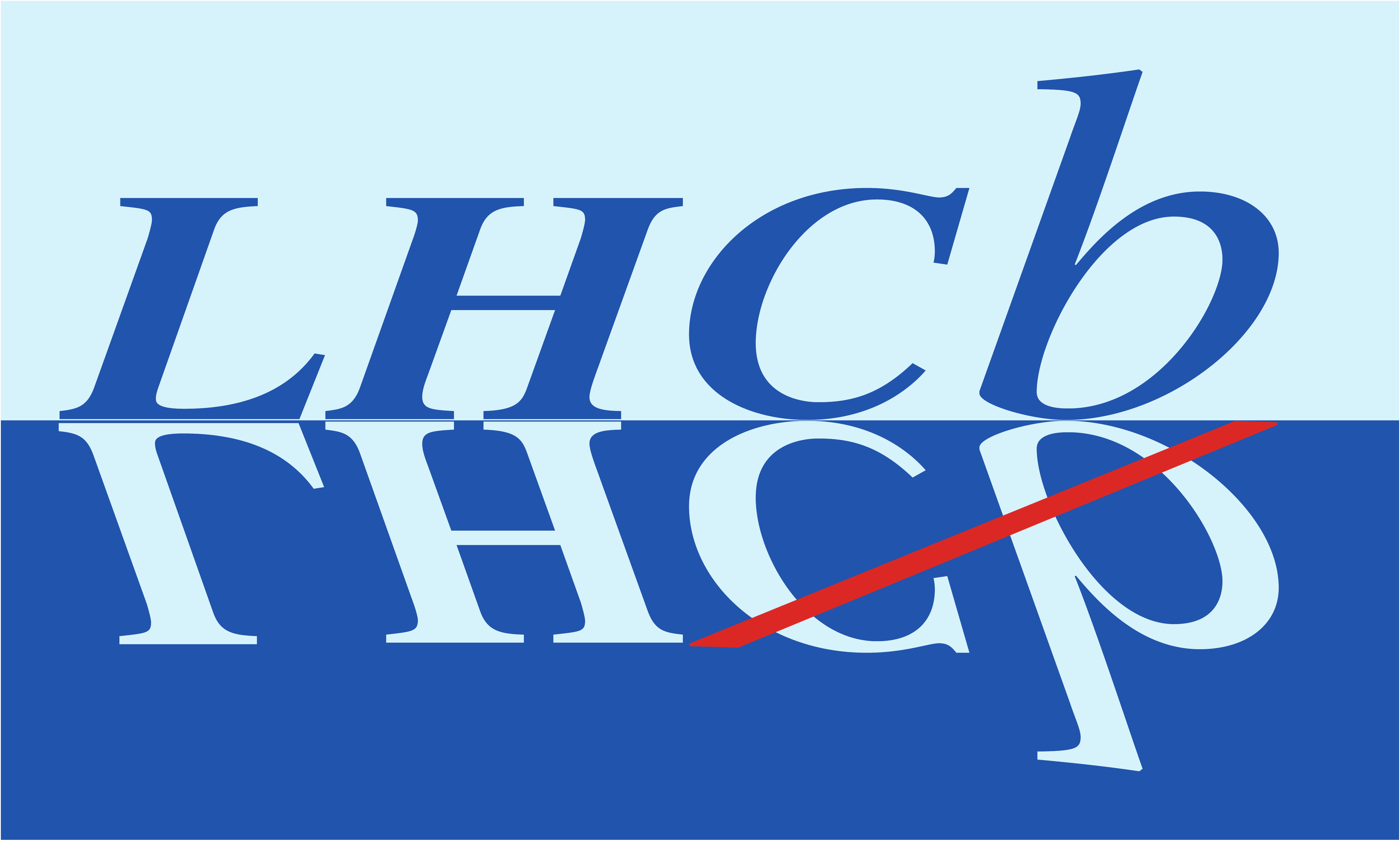}} & &}%
{\vspace*{-1.2cm}\mbox{\!\!\!\includegraphics[width=.12\textwidth]{lhcb-logo.eps}} & &}%
\\
 & & CERN-EP-2019-076 \\  
 & & LHCb-PAPER-2019-016 \\  
 & & November 29, 2019 \\ 
 & & \\
\end{tabular*}

\vspace*{3.0cm}

{\normalfont\bfseries\boldmath\huge
\begin{center}
  \papertitle 
\end{center}
}

\vspace*{1.0cm}

\begin{center}
\paperauthors
\end{center}

\vspace{\fill}

\begin{abstract}
\input{abstract}  
\end{abstract}

\vspace*{1.0cm}

\begin{center}
  Published in Phys. Rev. Lett. 123 (2019) 211801
\end{center}

\vspace{\fill}

{\footnotesize 
\centerline{\copyright~\papercopyright. \href{\paperlicenceurl}{\paperlicence}.}}
\vspace*{2mm}

\end{titlepage}


\newpage
\setcounter{page}{2}
\mbox{~}
%
%
%
%

\cleardoublepage

%% file: abstract.tex
Results are reported from a search for the rare decays \BsToTauMu and \BdToTauMu, where the tau lepton is reconstructed in the channel \tauppp.
These processes are effectively forbidden in the standard model, but they can potentially occur at detectable rates in models of new physics that can induce lepton-flavor-violating decays.
The search is based on a data sample corresponding to 3\invfb of proton-proton collisions recorded by the LHCb experiment in 2011 and 2012.
The event yields observed in the signal regions for both processes are consistent with the expected standard model backgrounds.
Because of the limited mass resolution arising from the undetected tau neutrino, the \Bs and \Bd signal regions are highly overlapping.
Assuming no contribution from \BdToTauMu, the upper limit  $\BR\left( \BsToTauMu\right) < 4.2 \times 10^{-5}$ is obtained at 95\% confidence level.
If no contribution from \BsToTauMu is assumed, a limit of $\BR\left( \BdToTauMu\right) <1.4 \times 10^{-5}$ is obtained at 95\% confidence level.
These results represent the first limit on $\BR\left( \BsToTauMu\right)$ and the most stringent limit on  $\BR\left( \BdToTauMu\right)$.

%% file: content.tex
Lepton-flavor-violating decays of mesons containing \bquark quarks, such as $\decay{\Bd(b\overline{d})}{\taupm\mump}$ and $\decay{\Bs(b\overline{s})}{\taupm\mump}$, are extremely suppressed in the Standard Model (SM), with expected branching fractions of order 10$^{-54}$~\cite{LFVsuppression}. (The inclusion of charge-conjugate processes is implied throughout this Letter.)
These processes involve not only quantum loops, but also neutrino oscillations.
Signals at the level expected in the SM lie far below current and foreseen experimental sensitivities.
However, many theoretical models proposed to explain possible experimental tensions observed in other \B-meson decays (discussed below) naturally allow for branching fractions that are within current sensitivity.
Among them, models containing a heavy neutral gauge boson ($\Z^{'}$) could lead to a \BsToTauMu branching fraction of up to $10^{-8}$ ~\cite{LFVbtos,1504.07928} when only left-handed or right-handed couplings to quarks are considered, or of the order of $10^{-6}$~\cite{1504.07928} if both are allowed.
In models with either scalar or vector leptoquarks, the largest predictions for the \BsToTauMu branching fraction range from $10^{-9}$ to $10^{-5}$, depending on the assumed leptoquark mass~\cite{1503.01084,1608.07583,1801.02895}.
The three-site Pati-Salam gauge model favours values for this branching fraction in the range 10$^{-4}$--10$^{-6}$~\cite{1805.09328,1903.11517}. 

The SM predicts that the electroweak couplings for the three lepton families are universal, a result referred to as Lepton Flavor Universality (LFU).
Experimental tests of LFU performed using $\bquark \rightarrow s \ell^+ \ell^-$ and $b\rightarrow c \ell^- \overline{\nu}$ processes show tensions with respect to the SM predictions for the observables $R_{\kaon^{(*)}}$~\cite{LHCb-PAPER-2017-013,LHCb-PAPER-2019-009b} and ${\cal R}(D^{(*)})$~\cite{HFLAV16}.
For the latter, the observed discrepancy with respect to the SM prediction is greater than 3 standard deviations.
Because theoretical models that can account for the possible LFU effects observed in data often predict Lepton Flavor Violation (LFV) as well~\cite{PhysRevLett.114.091801}, searches for LFV processes provide a powerful signature for probing these models.

An upper limit $\BR\left(\BdToTauMu\right)< 2.2\times 10^{-5} $ at $90\%$ confidence level (CL) was obtained by the \babar collaboration~\cite{babartaumu}. 
There are currently no experimental results for the \BsToTauMu mode.

This Letter reports results from the first search for the decay \BsToTauMu, along with the most stringent limit on the process \BdToTauMu.
The analysis is performed on data corresponding to an integrated  luminosity of $3\invfb$ of proton-proton ($pp$) collisions, recorded with the \lhcb detector during the years 2011 and 2012 at centre-of-mass energies of $7$ and $8\tev$, respectively.
The \tauon leptons are reconstructed through the decay \tauppp, which mainly proceeds via the production of two intermediate resonances, \mbox{$a_{1}(1260)^-\to \pi^+\pi^-\pi^-$} and \mbox{$\rho(770)^0\to \pi^+\pi^-$}~\cite{Schael:2005am}, which help in the signal selection.
In this mode, the \tauon decay vertex can be precisely reconstructed, facilitating a good reconstruction of the \B-meson invariant mass despite the undetected neutrino.
To avoid experimenter bias, the \B-meson invariant-mass signal region was not examined until the selection and fit procedures were finalised.
The signal yield is determined by performing an unbinned maximum-likelihood fit to the reconstructed \B-meson invariant-mass distribution and is converted into a branching fraction using the decay \BDpi as a normalisation channel.

The \lhcb detector~\cite{Alves:2008zz,LHCb-DP-2014-002} is a single-arm forward
spectrometer covering the \mbox{pseudorapidity} range $2<\eta <5$,
designed for the study of particles containing \bquark or \cquark
quarks. The detector includes a high-precision tracking system
consisting of a silicon-strip vertex detector surrounding the $pp$
interaction region, a large-area silicon-strip detector located
upstream of a dipole magnet with a bending power of about
$4{\mathrm{\,Tm}}$, and three stations of silicon-strip detectors and straw
drift tubes placed downstream of the magnet.
The tracking system provides a measurement of the momentum, \ptot, of charged particles with
a relative uncertainty varying from 0.5\% at low momentum to 1.0\% at 200\gevc.
The minimum distance of a track to a primary vertex (PV), the impact parameter (IP), 
is measured with a resolution of $(15+29/\pt)\mum$,
where \pt is the component of the momentum transverse to the beam, in\,\gevc.
Different types of charged hadrons are distinguished using information
from two ring-imaging Cherenkov detectors. 
Photons, electrons and hadrons are identified by a calorimeter system consisting of
scintillating-pad and preshower detectors, an electromagnetic and a hadronic calorimeter. Muons are identified by a
system composed of alternating layers of iron and multiwire
proportional chambers.

The on-line event selection is performed by a trigger~\cite{LHCb-DP-2012-004} consisting of a hardware stage based on information from the calorimeter and muon systems, followed by a software stage, which performs a full event reconstruction. 
At the hardware trigger stage, signal candidates are required to have a muon with high \pt, while, for the normalisation sample, events are required to have a hadron with high transverse energy in the calorimeters.
The software trigger requires a two-, three-, or four-track secondary vertex with a significant displacement from any primary $pp$ interaction vertex. 
A multivariate algorithm~\cite{BBDT} is used to identify secondary vertices consistent with the decay of a \bquark hadron.
At least one charged particle must have a transverse momentum $\pt > 1.0\: (1.6)\gevc$ for muons (hadrons), and must be inconsistent with originating from a PV.

Simulation is used to optimise the selection, determine the signal model for the fit and obtain the selection efficiencies.
In the simulation, $pp$ collisions are generated using \pythia~\cite{Sjostrand:2006za,*Sjostrand:2007gs} with a specific \lhcb configuration~\cite{LHCb-PROC-2010-056}.  
The $\tau$ decay is simulated using the \tauola decay library tuned with \babar data \cite{TauolaBabar}, while the
decays of all other unstable particles are described by \evtgen~\cite{Lange:2001uf}.
Final-state radiation is accounted for using \photos~\cite{Golonka:2005pn}. 
The interaction of the generated particles with the detector, and its response, are implemented using the \geant toolkit~\cite{Allison:2006ve, *Agostinelli:2002hh}, as described in Ref.~\cite{LHCb-PROC-2011-006}.

Both signal and normalisation candidates are formed using tracks that are inconsistent with originating from any PV.
Candidate \tauppp and \Dkpp decays are reconstructed from three tracks forming a good-quality vertex and with particle identification information corresponding to their assumed particle hypotheses.
Candidate \BdorsToTauMu decays are formed by combining a reconstructed \tauon lepton and an oppositely charged track identified as a muon.
A control sample of same-sign candidates, which are formed by a \tauon lepton and a muon with identical charges, is also selected to serve, during the selection process, as a proxy for the large component of the background in which the muon and the tau candidate charges are uncorrelated.
For the normalisation mode, \BToDpi candidates are made out of a reconstructed \D meson and an oppositely charged track identified as a pion.
The decay vertex of the signal or normalisation \B candidate is determined through a fit to all reconstructed particles in the decay chain~\cite{Hulsbergen:2005pu}, which is required to be of good quality.
The \B-meson \pt is required to be greater than $5\gevc$ for both signal and normalisation modes.

\begin{figure}
\centering
\includegraphics[width=0.49\textwidth]{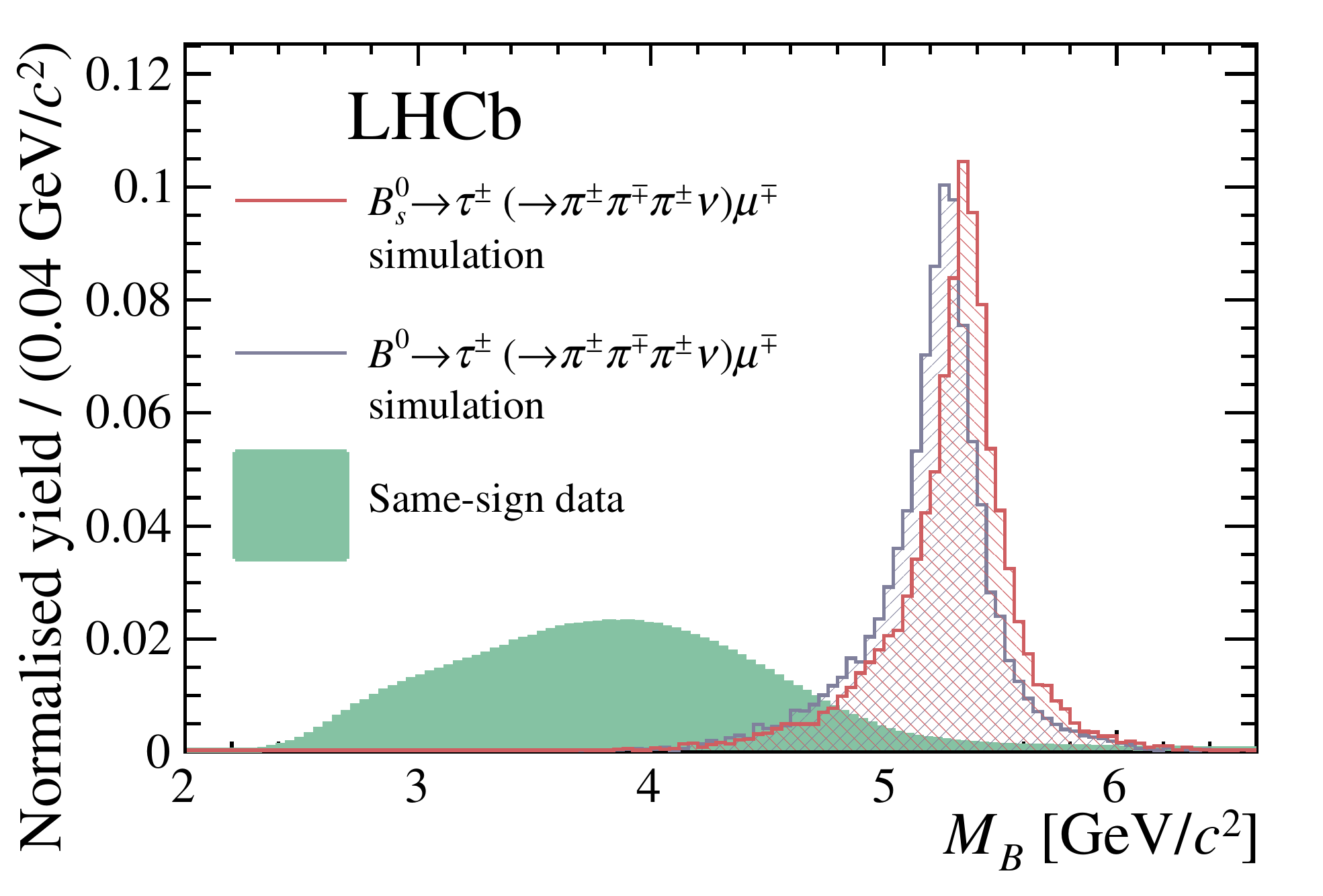}
\caption{Normalized distributions of the reconstructed invariant mass for \Bs and \Bd in simulated event samples and for same-sign candidates in data, after applying the initial event selection (see the text).}
\label{Fig:visible and analytical B masses}
\end{figure}

While the neutrino from the \tauon decay escapes detection,
its momentum vector can be constrained from the measured positions of the primary and \tauon decay vertices, the momenta of the muon and  the three pions, and the trajectory of the muon.
Then, by imposing the requirements that the mass of the system formed from the three pions and the unobserved neutrino corresponds to the mass of the tau lepton, and by requiring that the \B decay vertex lies on the trajectories of the muon, of the tau lepton, and of the \B meson, the invariant mass of the \BdorBs candidate can be determined analytically up to a twofold ambiguity.
Because of the quadratic nature of the equation, the computed masses may be unphysical.
This occurs in 32\% of the selected signal in the simulated event sample due to measurement resolutions and in 48\% of the same-sign candidates in data.
These candidates are removed, thereby improving the signal-to-background ratio.
The solution whose distribution shows the largest separation between signal and background is used as the reconstructed \B invariant mass, \MB, in the analysis.
The distributions of \MB for candidates satisfying the previously described initial selection in the simulated signal samples and in the opposite-sign control sample in the data are shown in Fig.~\ref{Fig:visible and analytical B masses}.

To reduce the data to a manageable level and focus on the rejection of the most difficult backgrounds, the low-mass region with $\MB<4\gevcc$ is discarded. The signal loss due to this requirement is negligible.

To further reduce the background, additional requirements, optimised with same-sign candidates and simulated samples, are applied to the selected \BdorsToTauMu decays.
Taking advantage of the resonant structure of the \tauppp decay, candidates with both combinations of oppositely charged pions with invariant-masses below $550\mevcc$ are removed.
Candidates with a three-pion invariant mass greater than $1.8\gevcc$ are discarded to veto the background contribution due to {\decay{\Dp}{\pip\pim\pip}} decays.

A set of isolation variables is used to reduce background from decays with additional reconstructed particles.
The first class of isolation variables exploits the presence of activity in the calorimeter to identify the contribution of neutral particles contained in a cone centred on the \B or \tauon flight directions.
The second class is based on the presence of additional tracks consistent with originating from the \B or \tauon decay vertices, or uses a multivariate classifier, trained on simulated data, to discriminate against candidates whose decay products are compatible with forming good-quality vertices with other tracks in the event.
These variables are combined using a Boosted Decision Tree (BDT)~\cite{Breiman}, trained on same-sign candidates and simulated \BsToTauMu decays. 
Candidates with a BDT output compatible with that of  background are discarded. A second BDT is used to reduce to a negligible level the contribution of combinatorial background, which extends over the whole mass range but dominates at higher masses. It uses variables related to vertex quality and reconstructed particle opening angles and is trained on samples of same-sign candidates with $\MB>6.2\gevcc$ and simulated \BsToTauMu decays. 

Some background processes, such as \decay{\BdorBs}{\subdecay{\Ddorsm}{\mun\neumb}\pip\pim\pip}, have \MB distributions peaking in the signal region.
In these decays, the three pions come from the \B decay vertex, and therefore the reconstructed \B and \tauon decay vertices are very close.
Discarding candidates with a reconstructed \tauon decay-time significance lower than $1.8$ reduces this type of background to a negligible level while keeping $\sim$75$\%$ of signal, according to studies performed on simulation.
All previously described selection criteria also suppress a possible contribution from the \decay{\Bd}{\aonem\mup\neum} mode, whose selection efficiency is 60 times lower than that of the signal.
Its rate is currently unmeasured, but, given that the largest known $b\to u $ semileptonic decay branching fractions are of the order of $10^{-4}$, its branching fraction is not expected to be much higher. Events from the decay \tauppppiz passing the selection are also included as signal.

The selection procedure retains 17\;746 candidates.
According to studies based on simulations, the remaining background is dominated by $\decay{\BdorBs}{\D_{(s)}^{(*)}\mu\nu\text{X}}$ decays.

The selection efficiencies for the signal and normalisation modes, $\epsilon_{\BdorBs\rightarrow\tauon\mu}$ and $\epsilon_{B\rightarrow D\pi}$, respectively, are estimated using simulation or, whenever possible, data. 
The efficiency $\epsilon_{\BdorBs\rightarrow\tauon\mu}$ includes those for both \tauppp and \tauppppiz decays, where the latter is weighted by the ratio of the two branching fractions.
The \tauppppiz channel contributes by $\sim16\%$ to the extracted signal yield.
The tracking and particle identification efficiencies are determined using data~\cite{LHCb-DP-2013-002,LHCb-PUB-2016-021}.
The trigger efficiency for the normalisation channel is estimated using a trigger-unbiased subsample made of events which have been triggered independently of the normalisation candidate.
For the signal, muons from \BToJpsiK decays are used to evaluate the muon trigger efficiency and corrections are applied to the simulated signal samples. To account for differences between the control and the signal samples, the efficiency is computed as a function of the muon \pt and IP. Simulation as well  as \BToJpsiK decays is used to determine the software-trigger efficiency and its systematic uncertainty.

The signal yield for the normalisation mode is obtained from a fit to the invariant-mass distribution of the \BToDpi candidates. In the fit the signal is modelled by the sum of two Crystal Ball (CB)~\cite{Skwarnicki:1986xj} functions, with tails on opposite sides, having common means and widths, but independent tail parameters. The tail parameters are fixed to values determined from a fit to a sample of \BDpi simulated decays, while all other parameters are left free. The small background contribution is described by an exponential function.
The measured yield of the \BToDpi mode is $N^{\text{norm}}=22\;588\pm176$ where the uncertainty is statistical only.

The \BdorsToTauMu branching fractions can be written as
\begin{equation}\label{eq:BR}
\BR\left( \BdorsToTauMu \right) = \alpha_{(s)}^{\text{norm}} \cdot N^{\text{sig}}_{(s)},
\end{equation}
where $N^{\text{sig}}_{(s)}$ is the number of observed \BdorsToTauMu decays and $\alpha_{(s)}^{\text{norm}}$ a normalisation factor.
The latter is defined by
\begin{equation}\label{eq:Normalization factor}
\alpha_{(s)}^{\text{norm}} = \frac{f_{\B^0}}{f_{\B^0_{(s)}}} \cdot \frac{\BR\left(\BDpi\right)}{\BR\left(\tauppp\right)} 
 \cdot \frac{\epsilon_{B\rightarrow D\pi}}{\epsilon_{\BdorBs\rightarrow\tauon\mu}} \cdot \frac{1}{N^{\text{norm}}},
\end{equation}
using externally measured quantities: the ratio of \mbox{\bquark-quark} hadronisation fractions to \Bs and \Bz mesons, $f_\Bs/f_\Bz=0.259 \pm 0.015$~\cite{fsfd}, $\BR\left(\BDpi\right)=(2.26 \pm 0.14)\times 10^{-4}$~\cite{PDG2018} and $\BR\left(\tauppp\right)=(9.02 \pm 0.05)\%$~\cite{PDG2018}.
The measured values of $\alpha_{(s)}^{\text{norm}}$ for the \Bs and \Bz modes are, respectively,
\begin{equation}\label{eq:alphas}
\begin{split}
\alpha_s^{\text{norm}} &= (4.32\pm0.19\pm0.45\pm0.36) \times 10^{-7} \text{ and}\\
\alpha^{\text{norm}}   &= (1.25\pm0.06\pm0.13\pm0.08) \times 10^{-7},
\end{split}
\end{equation}
where the three quoted uncertainties are the statistical uncertainty due to the sizes of the signal and normalisation simulated and data samples, the systematic uncertainty on the selection efficiencies (dominated by the trigger efficiency contribution, $\sim$11$\%$) and the total uncertainty on the externally measured quantities.

\begin{figure}
\centering
\includegraphics[width=0.49\textwidth]{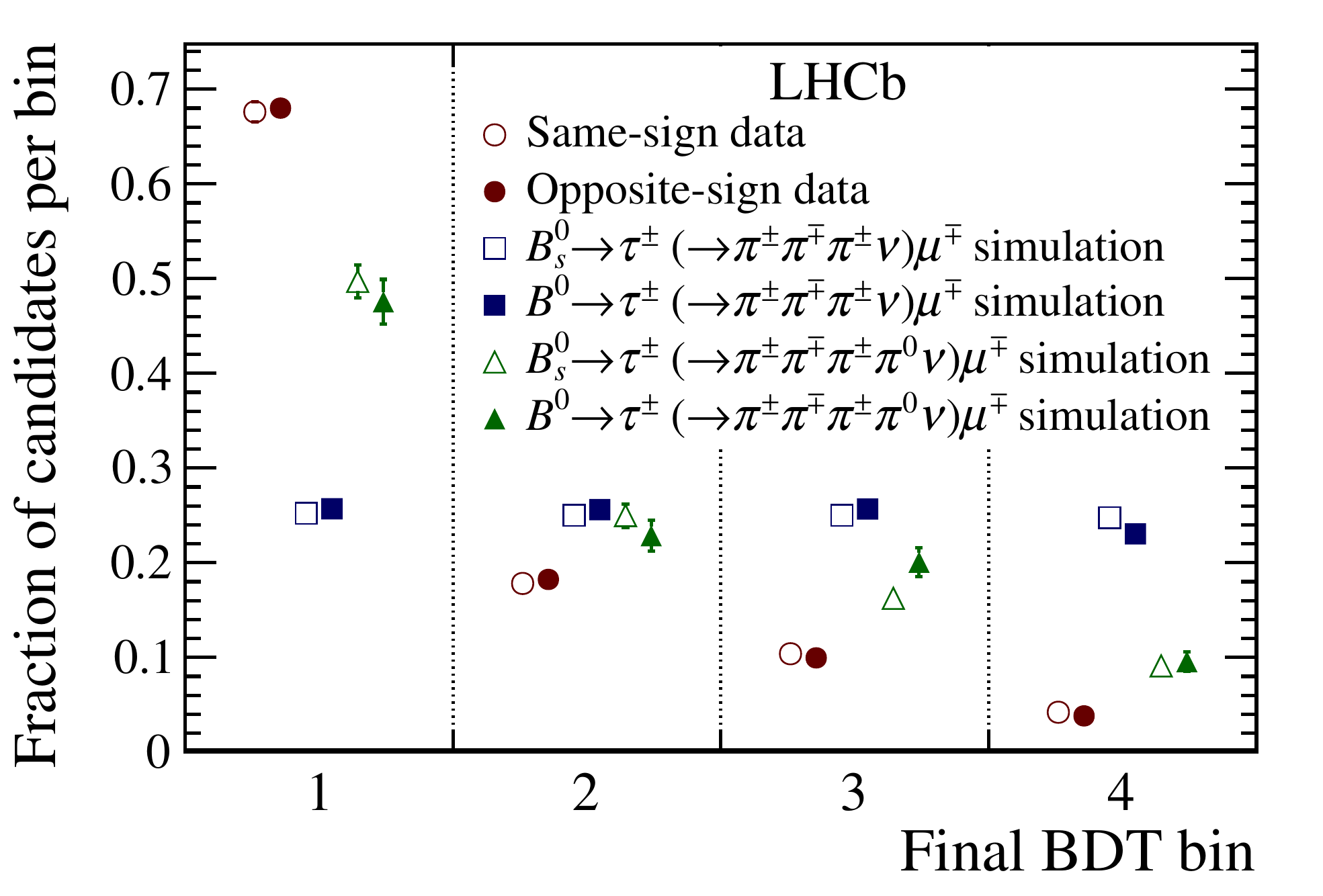}
\caption{Final BDT output binned distributions for data and simulated signal samples. The markers are displaced horizontally to improve visibility.}
\label{Fig:final BST output}
\end{figure}

A final BDT is built to split the selected candidates into four samples with different signal-to-background ratios. 
It combines 16 discriminating variables, none of which are correlated with the \B-meson invariant mass.
The most important ones are the invariant masses of the three-pion system and of the two combinations of oppositely charged pions, the \B-meson IP and flight distance significances, and the output of the BDT based on isolation variables.
The output of the BDT is transformed to have a uniform distribution between 0 and 1 for \Bstaumu simulated decays. As a consequence, its distribution for the background peaks at low BDT values. All samples are divided into four bins of equal width in BDT output. Their distributions are shown in Fig.~\ref{Fig:final BST output}. 

\begin{figure}
\begin{center}
    \includegraphics[width=.45\textwidth]{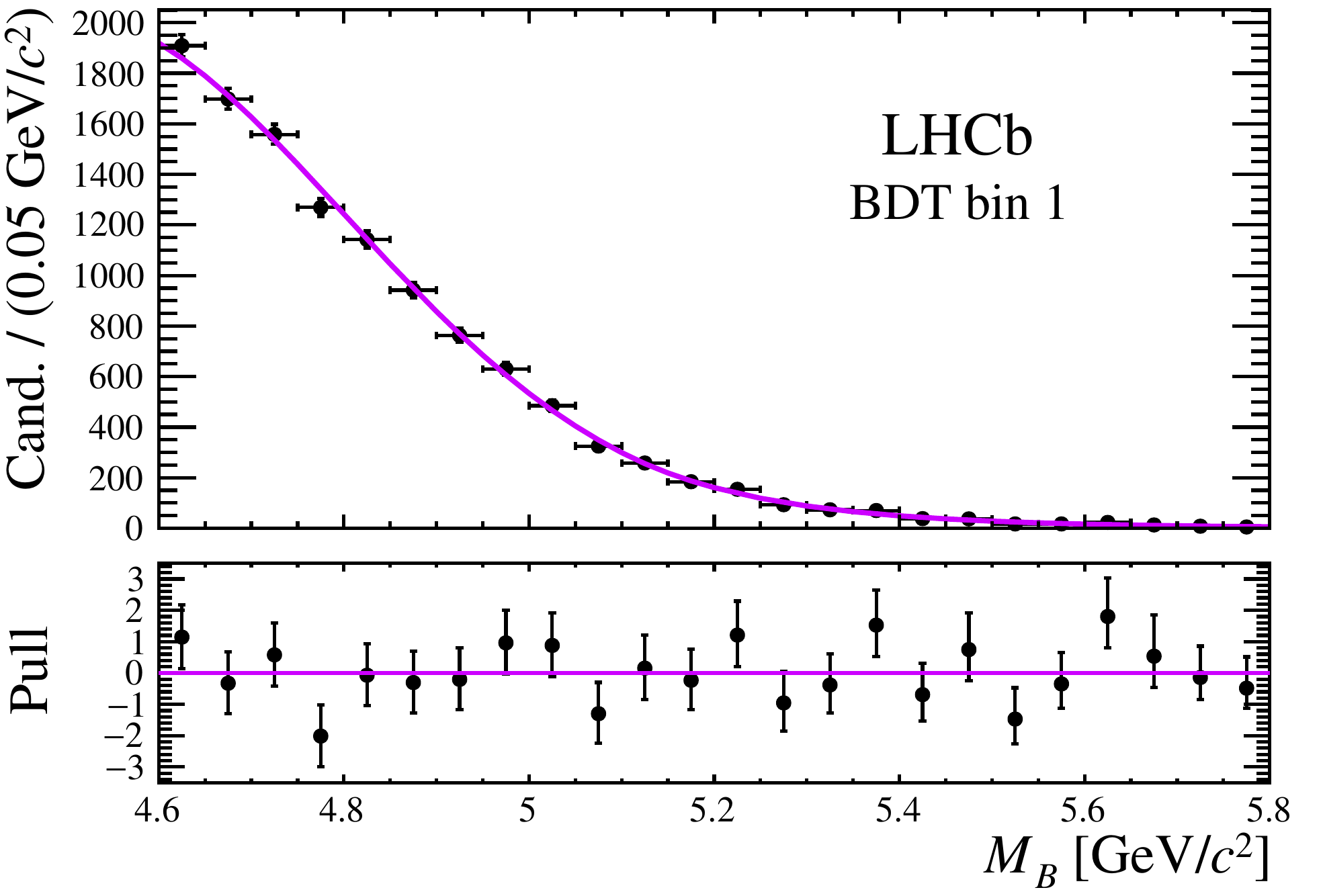}
    \includegraphics[width=.45\textwidth]{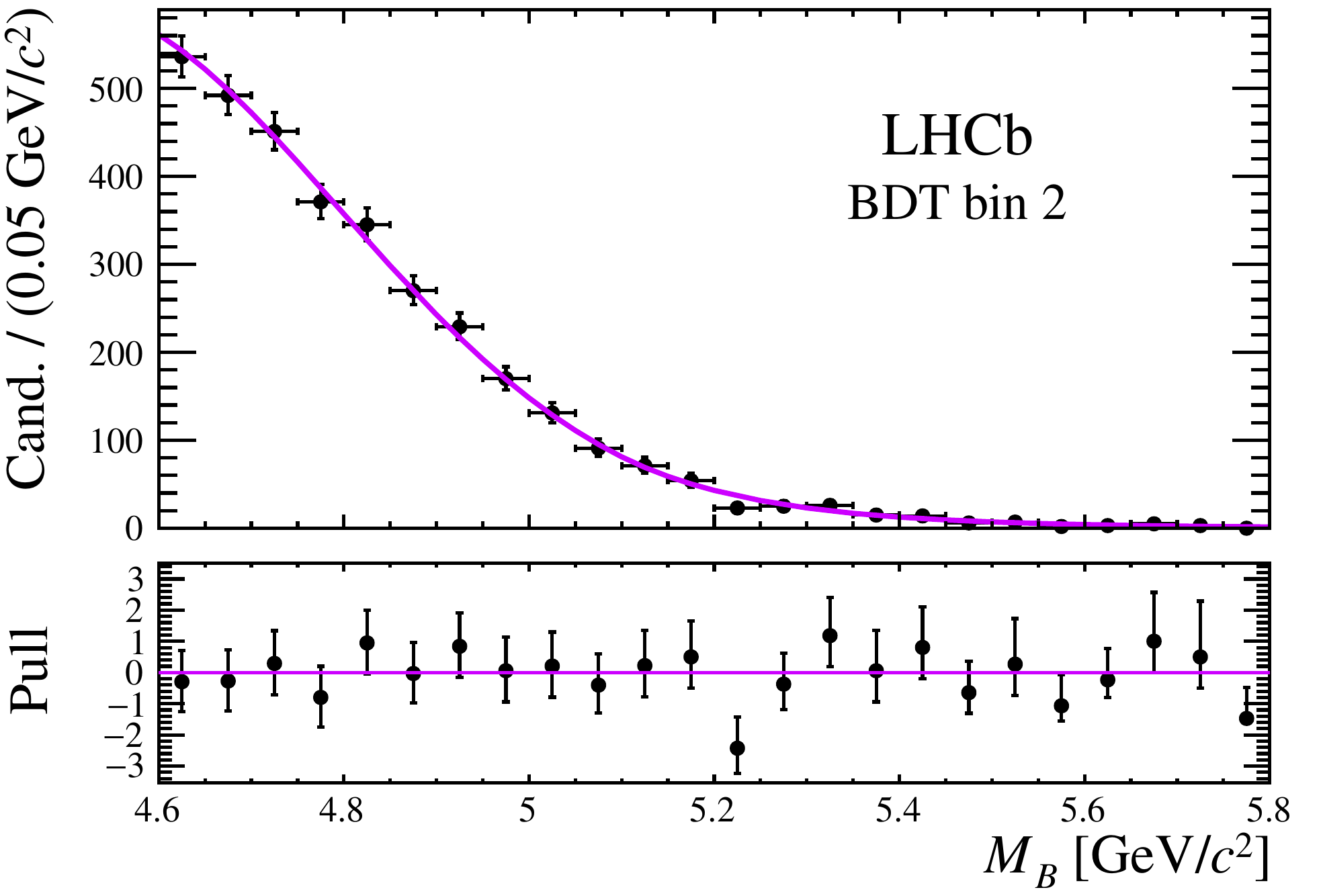}
    \includegraphics[width=.45\textwidth]{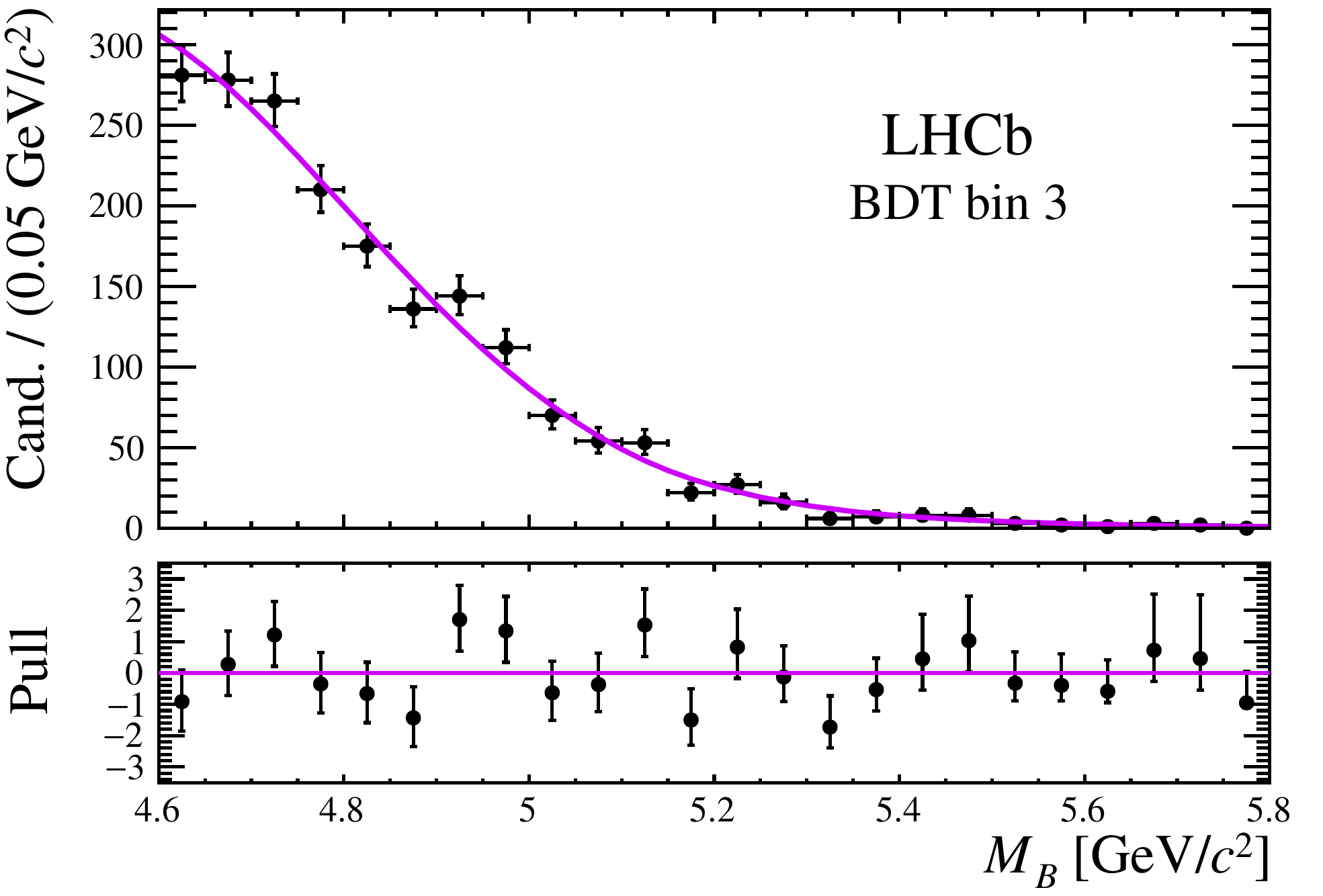}
    \includegraphics[width=.45\textwidth]{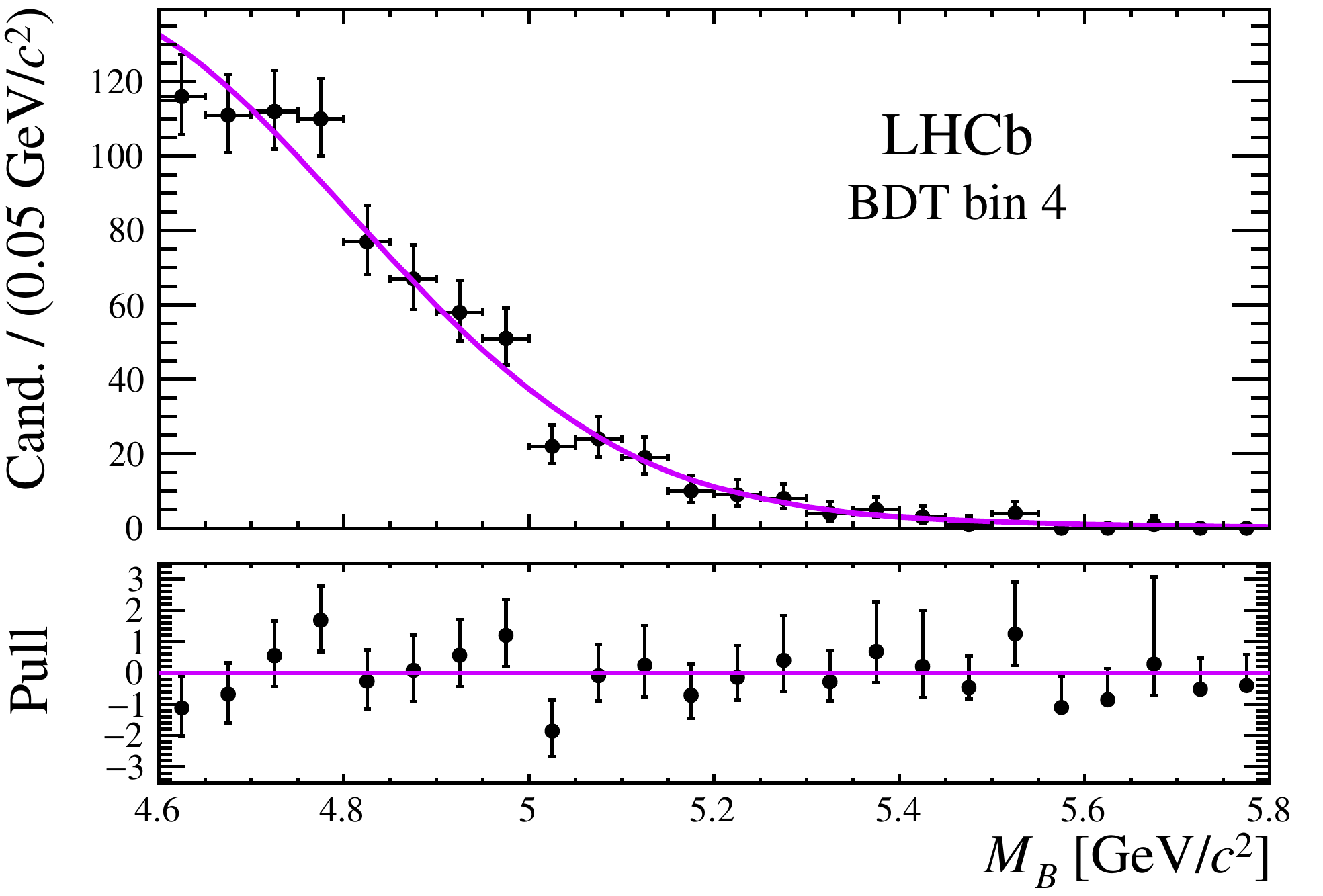}
\end{center}
\caption{Distributions of the reconstructed \B invariant-mass in data in the four final BDT bins with the projections of the fit for the \Bs signal-only hypothesis overlaid. The lower-part of each figure shows the normalised residuals.}
\label{Fig: Bds OS fit shared tails background parameters}
\end{figure}

The signal yield is evaluated by performing a simultaneous unbinned maximum-likelihood fit to the \MB distributions in the range $[4.6,5.8]\gevcc$ of the four samples corresponding to different BDT bins.
In each bin, the data are described by the sum of a signal and a background component.
The background shape is modelled by the upper tail of a reversed CB function, whose peak position and tail parameters are shared among BDT bins.
For the determination of the systematic uncertainties, different sets of constrained parameters or alternative background models, such as the sum of two Gaussian functions, are considered.
The signal shapes are described by double-sided Hypatia functions~\cite{doublesidedhypatia} whose parameters are initialized to the values obtained from a fit to the \BsToTauMu and \BdToTauMu simulated samples and allowed to vary within Gaussian constraints accounting for possible discrepancies between data and simulation.
The width of the Hypatia functions are $\sim330\mevcc$ for both signal modes in the most sensitive BDT bin.
As the separation between \BsToTauMu and \BdToTauMu signal shapes is limited, two independent fits are performed while assuming the contribution of either the \Bs or the \Bz signal only.
The signal fractional yields in each BDT bin are Gaussian constrained according to their expected values and uncertainties. 
The fit result corresponding to the hypothesis of the \Bs signal only is shown in Fig.~\ref{Fig: Bds OS fit shared tails background parameters}.
The fit procedure is validated by performing fits to a set of pseudoexperiments where the mass distributions are randomly generated according to the background model observed in the data.
The pulls of all fitted parameters are normally distributed except those of the signal yields $N^{\text{sig}}$, which have the expected widths but exhibit a very small bias of $-3\pm1$ ($2\pm2$) events for the \Bs (\Bd) mode.
This effect is accounted for by adding the bias to $N^{\text{sig}}$ in the simultaneous fits to the four BDT regions for both \Bd and \Bs.
The obtained signal yields are
\begin{equation*}
\begin{split}
N^\text{sig}_{\BsToTauMu} &= -16 \pm 38 \text{ and}\\
N^\text{sig}_{\BdToTauMu} &= -65 \pm 58, 
\end{split}
\end{equation*}
where the uncertainties account for the statistical ones as well as those on the signal and background shape parameters.
They show no evidence of any signal excess.

Using the calculated values of the normalisation factors $\alpha^{\text{norm}}$ and $\alpha_{s}^{\text{norm}}$ from Eq.~\ref{eq:BR} together with Eq.~\ref{eq:alphas}, the observed yields from the likelihood fits are translated into upper limits on the branching fractions using the CLs method~\cite{CLsMethod,AsymptoticCLs}.
The total uncertainty on the normalisation factor is accounted for as an additional Gaussian constraint in the simultaneous fit.
Furthermore, a systematic uncertainty on the signal yield of 34 (41) for the \Bs (\Bz) mode, derived using different sets of constrained parameters or alternative background models, is added to account for the uncertainties in the background shape.
The expected and observed CLs values as a function of the branching fraction are shown in the Supplemental Material~\cite{supplemental}.
The corresponding limits on the \Bs and \Bd branching fractions at $90\%$ and $95\%$ CL are given in Table~\ref{tab:Final limits} assuming negligible contribution from the \decay{\Bd}{\aonem\mup\neum} mode. A possible residual contribution of this background would lower the expected limits by $\sim$16$\%\times (\BR(\decay{\Bd}{\aonem\mup\neum} )/10^{-4})$. 
{\setlength{\extrarowheight}{2pt}%
\begin{table}
\centering
\caption{Expected and observed $90\%$ and $95\%$ CL limits on the \BdorsToTauMu\ branching fraction.\label{tab:Final limits}}
\begin{tabular}{c|c|c|c}
Mode         & Limit    & $90\%$ CL           & $95\%$ CL           \\\hline\hline
\BsToTauMu   & Observed & $3.4\times 10^{-5}$ & $4.2\times 10^{-5}$ \\
             & Expected & $3.9\times 10^{-5}$ & $4.7\times 10^{-5}$ \\\hline
\BdToTauMu   & Observed & $1.2\times 10^{-5}$ & $1.4\times 10^{-5}$ \\
             & Expected & $1.6\times 10^{-5}$ & $1.9\times 10^{-5}$ \\
\end{tabular}
\end{table} 
}
The impact of systematic uncertainties on the final limits is about 35$\%$, dominated by the uncertainty on the background model.

These results represent the best upper limits to date.
They constitute a factor $\sim$2 improvement with respect to the \babar result for the \Bd mode~\cite{babartaumu} and the first measurement for the \Bs mode.
The allowed range on the \BsToTauMu branching fraction preferred by the three-site Pati-Salam model~\cite{1805.09328,1903.11517} is significantly reduced by the results presented in this Letter.

%% file: acknowledgements.tex
\section*{Acknowledgements}
%
%
\noindent We express our gratitude to our colleagues in the CERN
accelerator departments for the excellent performance of the LHC. We
thank the technical and administrative staff at the LHCb
institutes.
We acknowledge support from CERN and from the national agencies:
CAPES, CNPq, FAPERJ and FINEP (Brazil); 
MOST and NSFC (China); 
CNRS/IN2P3 (France); 
BMBF, DFG and MPG (Germany); 
INFN (Italy); 
NWO (Netherlands); 
MNiSW and NCN (Poland); 
MEN/IFA (Romania); 
MSHE (Russia); 
MinECo (Spain); 
SNSF and SER (Switzerland); 
NASU (Ukraine); 
STFC (United Kingdom); 
DOE NP and NSF (USA).
We acknowledge the computing resources that are provided by CERN, IN2P3
(France), KIT and DESY (Germany), INFN (Italy), SURF (Netherlands),
PIC (Spain), GridPP (United Kingdom), RRCKI and Yandex
LLC (Russia), CSCS (Switzerland), IFIN-HH (Romania), CBPF (Brazil),
PL-GRID (Poland) and OSC (USA).
We are indebted to the communities behind the multiple open-source
software packages on which we depend.
Individual groups or members have received support from
AvH Foundation (Germany);
EPLANET, Marie Sk\l{}odowska-Curie Actions and ERC (European Union);
ANR, Labex P2IO and OCEVU, and R\'{e}gion Auvergne-Rh\^{o}ne-Alpes (France);
Key Research Program of Frontier Sciences of CAS, CAS PIFI, and the Thousand Talents Program (China);
RFBR, RSF and Yandex LLC (Russia);
GVA, XuntaGal and GENCAT (Spain);
the Royal Society
and the Leverhulme Trust (United Kingdom).

%% file: supplemental.tex
\clearpage

\section*{Supplemental material for LHCb-PAPER-2019-016}
\label{sec:Supplemental}
\begin{figure}[h]
\begin{center}
\includegraphics[width=0.60\textwidth]{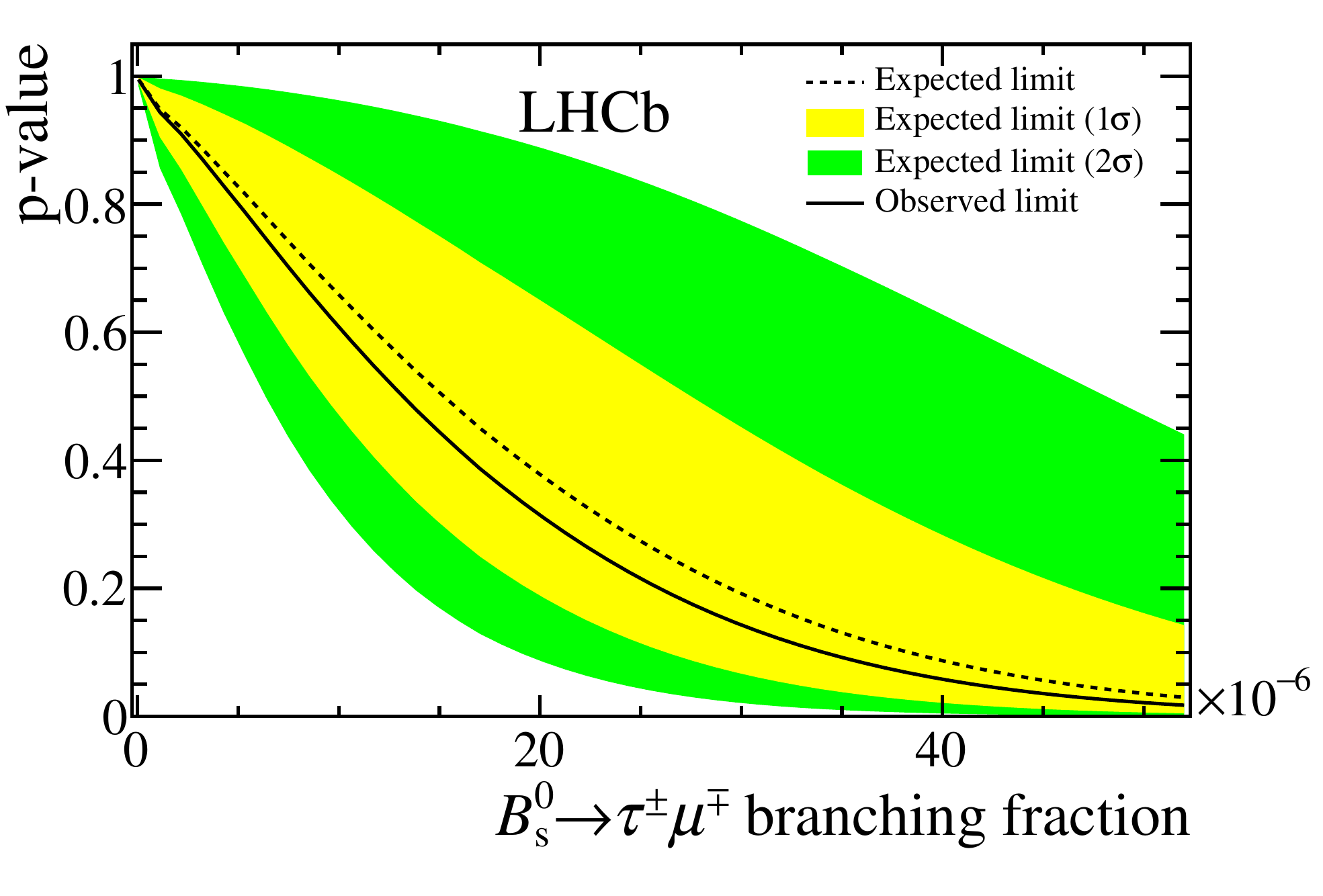}
\end{center}
\caption{The expected and observed p-values derived with the CLs method as a function of the \BsToTauMu branching fraction. }
\label{Fig:Brazillian Bs}
\end{figure}

\begin{figure}[h]
\begin{center}
\includegraphics[width=0.60\textwidth]{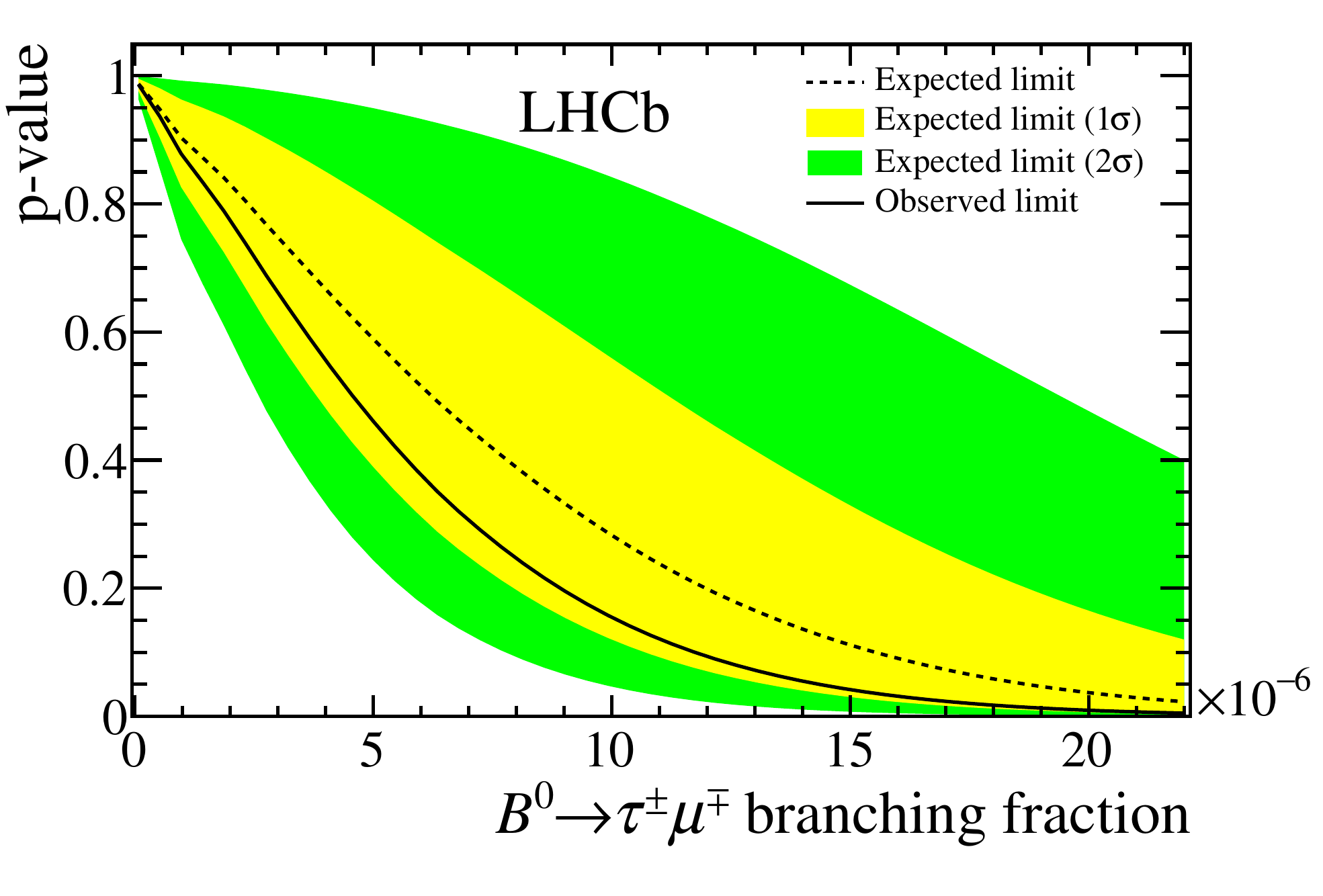}
\end{center}
\caption{The expected and observed p-values derived with the CLs method as a function of the \BdToTauMu branching fraction. }
\label{Fig:Brazillian Bd}
\end{figure}

Figures~\ref{Fig:Brazillian Bs} and~\ref{Fig:Brazillian Bd} show the expected and observed CLs values as a function of the \BsToTauMu and \BdToTauMu branching fractions.

%% file: LHCb_Authorship_10-Mar-2019.tex
\centerline
{\large\bf LHCb collaboration}
\begin
{flushleft}
\small
R.~Aaij$^{30}$,
C.~Abell{\'a}n~Beteta$^{47}$,
B.~Adeva$^{44}$,
M.~Adinolfi$^{51}$,
C.A.~Aidala$^{78}$,
Z.~Ajaltouni$^{8}$,
S.~Akar$^{62}$,
P.~Albicocco$^{21}$,
J.~Albrecht$^{13}$,
F.~Alessio$^{45}$
% ...

{\footnotesize \it

% ...
$ ^{8}$Universit{\'e} Clermont Auvergne, CNRS/IN2P3, LPC, Clermont-Ferrand, France\\
% ...
$ ^{13}$Fakult{\"a}t Physik, Technische Universit{\"a}t Dortmund, Dortmund, Germany\\
% ...
$ ^{21}$INFN Laboratori Nazionali di Frascati, Frascati, Italy\\
% ...
$ ^{30}$Nikhef National Institute for Subatomic Physics, Amsterdam, Netherlands\\
% ...
$ ^{44}$Instituto Galego de F{\'\i}sica de Altas Enerx{\'\i}as (IGFAE), Universidade de Santiago de Compostela, Santiago de Compostela, Spain\\
$ ^{45}$European Organization for Nuclear Research (CERN), Geneva, Switzerland\\
% ...
$ ^{47}$Physik-Institut, Universit{\"a}t Z{\"u}rich, Z{\"u}rich, Switzerland\\
% ...
$ ^{51}$H.H. Wills Physics Laboratory, University of Bristol, Bristol, United Kingdom\\
% ...
$ ^{62}$University of Cincinnati, Cincinnati, OH, United States\\
% ...
$ ^{65}$Syracuse University, Syracuse, NY, United States\\
% ...
$ ^{78}$University of Michigan, Ann Arbor, United States, associated to $^{65}$\\
\bigskip
% ...
}
\end{flushleft}